\documentclass[nofootinbib,notitlepage,superscriptaddress,twocolumn]{revtex4-2}
\pdfoutput=1

\usepackage{aas_macros,amsmath,amssymb,esint,graphicx,overpic,xcolor,mathrsfs,stackengine,array}
\usepackage[bottom]{footmisc}
\usepackage[colorlinks=true]{hyperref}
\definecolor{red}{RGB}{228,26,28}
\definecolor{green}{RGB}{77,175,74}
\definecolor{blue}{RGB}{55,126,184}
\definecolor{purple}{RGB}{152,78,163}

\let\originalleft\left
\let\originalright\right
\renewcommand{\left}{\mathopen{}\mathclose\bgroup\originalleft}
\renewcommand{\right}{\aftergroup\egroup\originalright}

\newcommand{\ab}[1]{\left|#1\right|}

\newcommand{\cu}[1]{\left\{#1\right\}}
\newcommand{\pa}[1]{\left(#1\right)}

\newcommand{\floor}[1]{\lfloor#1\rfloor}

\newcolumntype{P}[1]{>{\raggedright\arraybackslash}p{#1}}

\usepackage{comment}

\begin{document}

\title{Black Holes with Accretion Disks as High-Energy Particle Colliders}

\author{Delilah E.~A. Gates}
\email{delilah.gates@cfa.harvard.edu}
\affiliation{Center for Astrophysics $\arrowvert$ Harvard \& Smithsonian, 60 Garden Street, Cambridge, MA 02138, USA}
\affiliation{Black Hole Initiative at Harvard University, 20 Garden Street, Cambridge, MA 02138, USA}

\begin{abstract}
In principle, rapidly rotating black holes (BHs) with accretion disks---either prograde or retrograde---could naturally act as high-energy particle colliders (HEPCs) because particles falling in from infinity that collide with particles in the disk near the BH horizon can exhibit center-of-mass (CM) energies beyond the capabilities of terrestrial instruments. However, we show that scattering can moderate the CM energy of near-horizon collisions.
We find that rapidly rotating BHs with prograde accretion disks are the most viable candidates for astrophysical HEPCs,
whereas BHs with retrograde disks are disfavored, when scattering is incorporated. 
\end{abstract}
\maketitle

Over a decade and a half ago, the discovery that a pair of particles can collide near a rapidly rotating black hole (BH) with a large---formally divergent---center-of-mass (CM) energy introduced the prospect of BHs acting as astrophysical high-energy particle colliders (HEPCs), producing collision energies beyond the capabilities of terrestrial instruments~\cite{Banados2009}. 
This phenomenon, the Ba{\~n}ados--Silk--West (BSW) effect, can occur when one particle has angular momentum tuned to a critical value, while the other remains non-critical~\cite{Harada2014,Gates2025b}. 

Criticisms note the high spin of the BH, the fine-tuning of the critical particle, and the slow growth of the CM energy as the collision radius nears the horizon~\cite{Berti2009,Jacobson2010}. 
X-ray spectroscopic measurements, which indicate BHs can rapidly rotate, address the first criticism (e.g., ~\cite{Brenneman2013,Reynolds2021,Bambi2021,Draghis2023,Ricarte2025,Gates2025a}).
The fact that both non-critical and critical particles can fall from infinity or arise within an accretion flow (in particular the optically thick, geometrically thin accretion disks as described in the treatment by Cunningham~\cite{Cunningham1975}), addresses the second. Thus, prior works have argued that BHs with accretion disks may naturally act as HEPCs~\cite{Harada2011a,Harada2011b,Mummery2025,Gates2025b}.

In this work, we provide the first-ever address of the latter criticism by considering scattering, which can moderate the CM energy by limiting how close collisions occur to the horizon. 
We quantify the scattering-moderated CM energy in the presence of an accretion disk (prograde or retrograde) and identify the most astrophysically viable disk configuration for which a BH can act as an HEPC.
We also examine the BH without a disk (the original BSW setup) for comparison. 
Throughout, we adopt the following abbreviations: unbound (UB), marginally bound (MB), marginally stable (MS), spherical orbit (SO), circular orbit (CO), near-horizon (NH), prograde disk (PD), retrograde disk (RD), and no disk (ND). We work in units $G_N=c=1$.

\emph{BSW effect ---}
In the Kerr geometry, particle trajectories are characterized by conserved quantities $\cu{m,E,L,Q}$: mass, energy, angular momentum about the spin axis, and Carter constant, respectively~\cite{Carter1968}. 
The CM energy of two colliding particles is given by ${E}_{\rm cm}^2=m_{[1]}^2+m_{[\rm 2]}^2-2p_{[1]}\cdot p_{[2]}$. Through
\begin{align}
    -\frac{p_{[1]}\cdot p_{[2]}}{E_{[\rm 1]}E_{[\rm 2]}}=\frac{{\mathcal{F}_{\rm cm}(\theta)+r^2\mathcal{E}_{\rm cm}^2(r)}}{r^2+a^2\cos^2\theta},
\end{align}
we define the functions $\mathcal{F}_{\rm cm}$ and $\mathcal{E}_{\rm cm}$ which depend only on the dimensionless, energy-normalized quantities
\begin{align}
    \bar m=\frac{m}{E},\quad\lambda=\frac{L}{ME},\quad\eta=\frac{Q}{M^2E^2},
\end{align}
of both particles. 

In the flat-space limit ($r\gg M$),
\begin{align}
{E}_{\rm cm}^2\approx&m_{[1]}^2+m_{[\rm 2]}^2+ 2E_{[\rm 1]}E_{[\rm 2]} \mathcal{E}_{\rm cm}^2,\\
\mathcal{E}_{\rm cm}^2\approx& {1-\sigma\sqrt{1-\bar{m}^2_{[1]}}\sqrt{1-\bar{m}^2_{[2]}}},\\
 \sigma=&{\rm sign}\pa{{p^r_{[1]}p^r_{[2]}}}.
\end{align}
In this limit, $\mathcal{E}_{\rm cm}\in[0,\sqrt{2}]$. At finite radius, $\mathcal{E}_{\rm cm}$ may exceed this bound, serving as an ``enhancement factor'' that encodes general relativistic effects.
This factor diverges in the near-horizon limit ($r- r_{\rm H}\ll M$) if one of the particles has angular momentum tuned to the critical value $ \lambda=\Omega_{H}^{-1}$, the inverse of the angular velocity of the horizon, 
while the other remains generic (i.e., non-critical). 
$\mathcal{F}_{\rm cm}$ is always finite.

Near-extremal BHs admit a set of prograde SOs at radii that approach the extremal horizon $r_{\rm H}=M$ and become critical in the extremal limit~\cite{Teo2021}. Hence, critical particles may orbit at or spiral onto/out of these NHSO radii~\cite{Harada2014,Gates2025b}. 
The NHSOs include the prograde MSSOs, MBSOs, and UBSOs.
All retrograde SOs remain well separated from the horizon at extremality.

\emph{Collision particles ---}
We restrict colliding particles to two origins: infall from infinity 
(a proxy for originating from another distant astrophysical source) 
or motion within an accretion flow. 
Thus, the generic particle can be a MB/UB particle or an RD particle; whereas, the critical particle can be a MB/UB particle or a PD particle.
This gives three scenarios as shown in Tab.~\ref{tab:scenarios}.
The BH with a PD (BH--PD) permits scenarios 1 and 3; the BH with an RD (BH--RD) permits scenarios 2 and 3; and the BH with ND (BH--ND) permits only scenario 3.
\begin{table}
    \centering
 \begin{tabular}{|c|| c| c| c| c|} 
 \hline
 \Centerstack{{collision} {scenario}} & \Centerstack{{generic} {particle}}& \Centerstack{{critical} {particle}} & \Centerstack{{accretion-disk} {configuration(s)}} \\[1ex] 
 \hline
 \hline
 1 & {MB/UB} & {PD} & PD\\[1ex] 
 \hline
 2 & {RD} & {MB/UB} & RD\\[1ex] 
 \hline
 3 & {MB/UB} & {MB/UB} & RD,\ PD,\ ND\\[1ex] 
 \hline
 \end{tabular}
    \caption{The near-horizon, high-energy collision scenarios under the restriction that the particles originate either at infinity or in a Cunningham accretion disk permitted for an extremal BH with prograde disk, retrograde disk, and no disk.}
    \label{tab:scenarios}
\end{table}

\begin{figure}
    \centering
    \includegraphics[width=\linewidth]{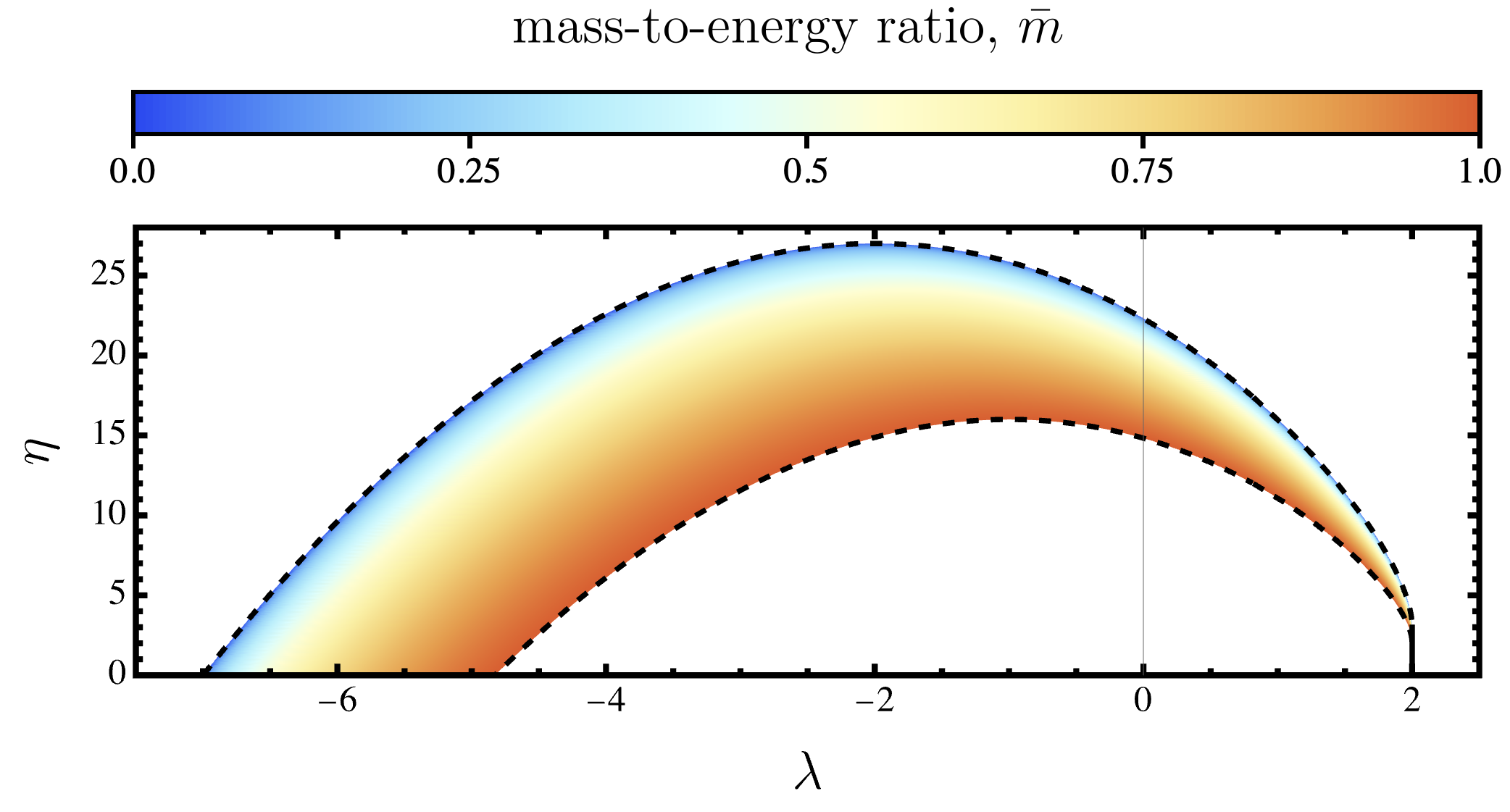}
    \caption{Conserved quantities, angular momentum about the spin axis $\lambda$ and Carter constant $\eta$, of marginally bound and unbound spherical orbiters at extremality. At fixed mass-to-energy ratio $\bar{m}$, $\eta(\lambda)$ is singular at $\lambda=2$, corresponding to near-horizon spherical orbits which are scaled to the horizon in the extremal limit.}
    \label{fig:ConsQuants}
\end{figure}

Disk particles orbit at fixed radii down to the prograde/retrograde MSCO radii $r_{\mathrm{msco}}^\pm$, with conserved quantities~\cite{Bardeen1972} 
\begin{align}
   \bar{m}_{\rm co}^{\pm}(r),\quad \lambda_{\rm co}^\pm (r), \quad \eta_{\rm co}^{\pm}=0.
\end{align}
Inside the MSCO radius, disk particles maintain the conserved quantities of the MSCO orbiter.

MB/UB particles have $\bar{m}\leq1$.
MB/UB generic particles that reach the horizon have angular momentum and Carter constant bounded by those of the spherical orbiters of the same mass-to-energy ratio~\cite{Compere2021,Wang2022}
\begin{align}
    \lambda_{\rm [g]}\in(\ \lambda_{\rm co}^-(\bar{m}),\lambda_{\rm co}^+(\bar{m})\ ), \quad \eta_{\rm [g]}\in[\ 0, \eta_{\rm so}(\lambda,\bar{m})\ ),
\end{align}
where the subscript ``$[\rm g]$'' denotes the conserved quantities of the generic particle. 
We plot $\eta_{\rm so}(\lambda,\bar{m})$ at extremality in Fig.~\ref{fig:ConsQuants}. 
MB/UB critical particles have the
conserved quantities of the NHSO orbiters, which in the extremal limit are~{\cite{Wilkins1972,Teo2021}}
\begin{align}
\lambda_{\rm [c]}=2,\quad \eta_{\rm [c]}\in[0,3-\bar{m}^2],
\end{align}
where the subscript ``$[\rm c]$'' denotes the conserved quantities of the critical particle.

\emph{Scattering-moderated CM energy ---}
A particle is unlikely to avoid scattering (i.e., remain along its geodesics) if it encounters the region of motion of other particles many times. Thus, we examine the CM energy bounds imposed by limiting the number of such encounters the particles undergo. 
The CM energy is bounded when an encounter must occur at a finite distance from the horizon, and diverges when an encounter can occur at the horizon.

The poloidal motion of a particle with initial coordinates $\cu{\theta_i,r_i}$ is described by functions $\theta(\tau,\theta_i)$ and $r(\tau,r_i)$, parametrized by Mino time $\tau$.
Particles with $\eta=0$ are confined to the equatorial plane. 
Particles with $\eta>0$ exhibit periodic libratory motion confined to $|\cos\theta|\le|\cos\theta_*|$, where $\theta_*\in[0,\pi/2]$, completing a half orbit in $\tau=\hat G_\theta$~\cite{Mino2003,Wang2022}. 
A particle crosses the equatorial plane when $\cos\theta(\tau,\theta_i)/\cos\theta_*=0$~\footnote{The crossing locations are well defined in the $\eta\to0$ limit.}.
The minimal number of crossings in elapsed time $\tau$ is $N=\floor{\tau/\hat G_\theta}$~\footnote{Crossings occur precisely at $\tau=N\hat G_\theta$ when $\theta_i=\pi/2$.}.
A particle that reaches the horizon in finite time $\tau=I_r^{\rm H}$ has minimal crossing number $N_{\rm H}=\floor{I_r^{\rm H}/\hat G_\theta}$. 
For an inward-directed particle ($\partial_\tau r(\tau,r_i)<0$), the maximal and minimal radii at a given crossing number $N$ are related by
\begin{align}
    \label{eq:RadCrossMin}
    \hat r_{{\rm max}}(N,r_i)&=r(\tau=N\hat G_\theta,r_i)=\hat r_{{\rm min}}(N-1,r_i).
\end{align}
For the MB/UB particles, $r_i=\infty$.

\begin{figure}
    \centering
    \includegraphics[width=\linewidth]{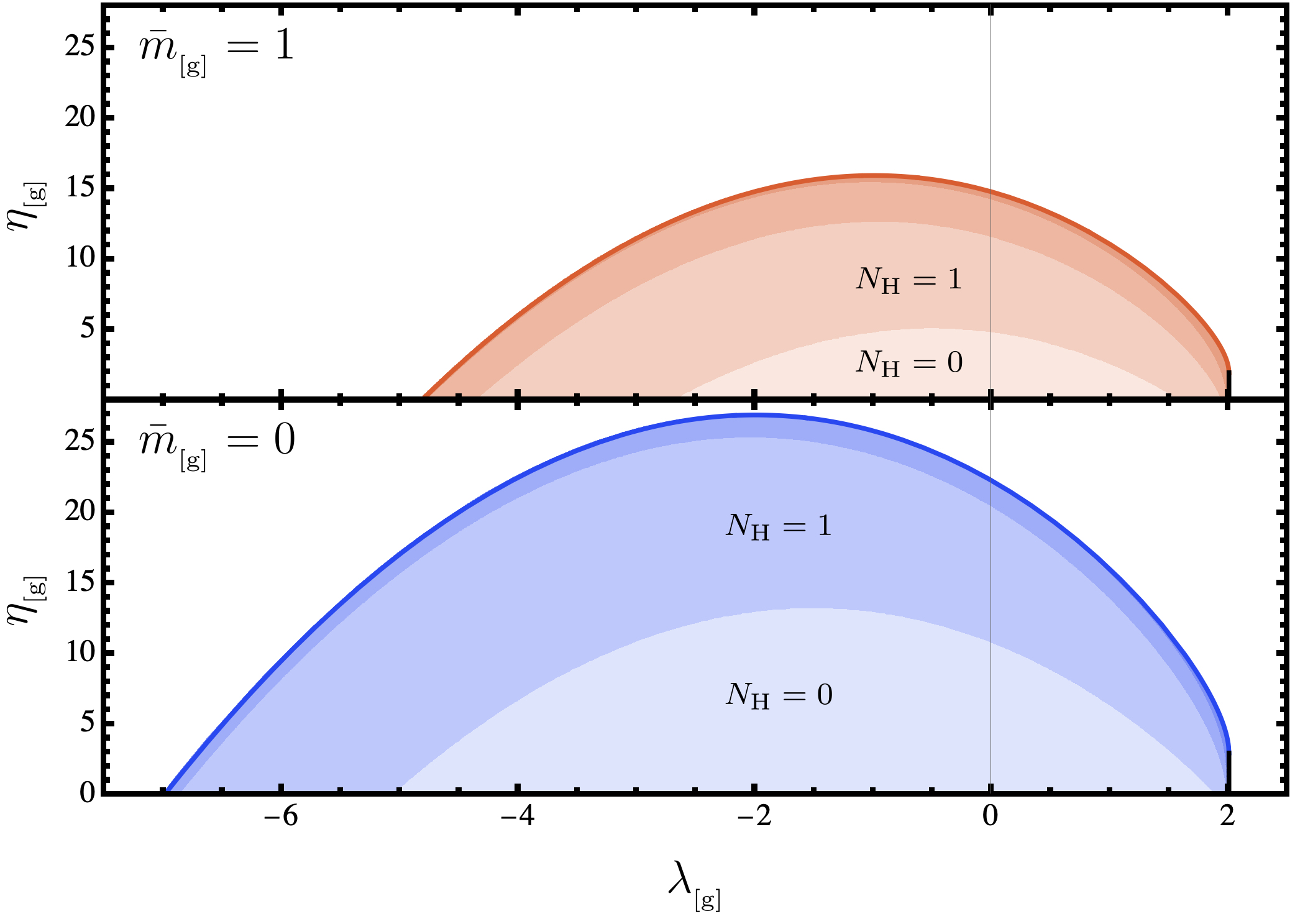}
    \caption{Minimal equatorial crossings that marginally bound (top) and null (bottom) particles undergo before reaching the horizon, $N_{\rm H}$, with the first two regions explicitly labeled. Regions of increasing crossing number are packed increasingly close to the boundary defined by the spherical orbits.}
    \label{fig:GenericCross}
\end{figure}

\begin{figure*}
    \centering
    \resizebox{\linewidth}{!}{\begin{tabular}{cc}
       \includegraphics[width=.45\linewidth]{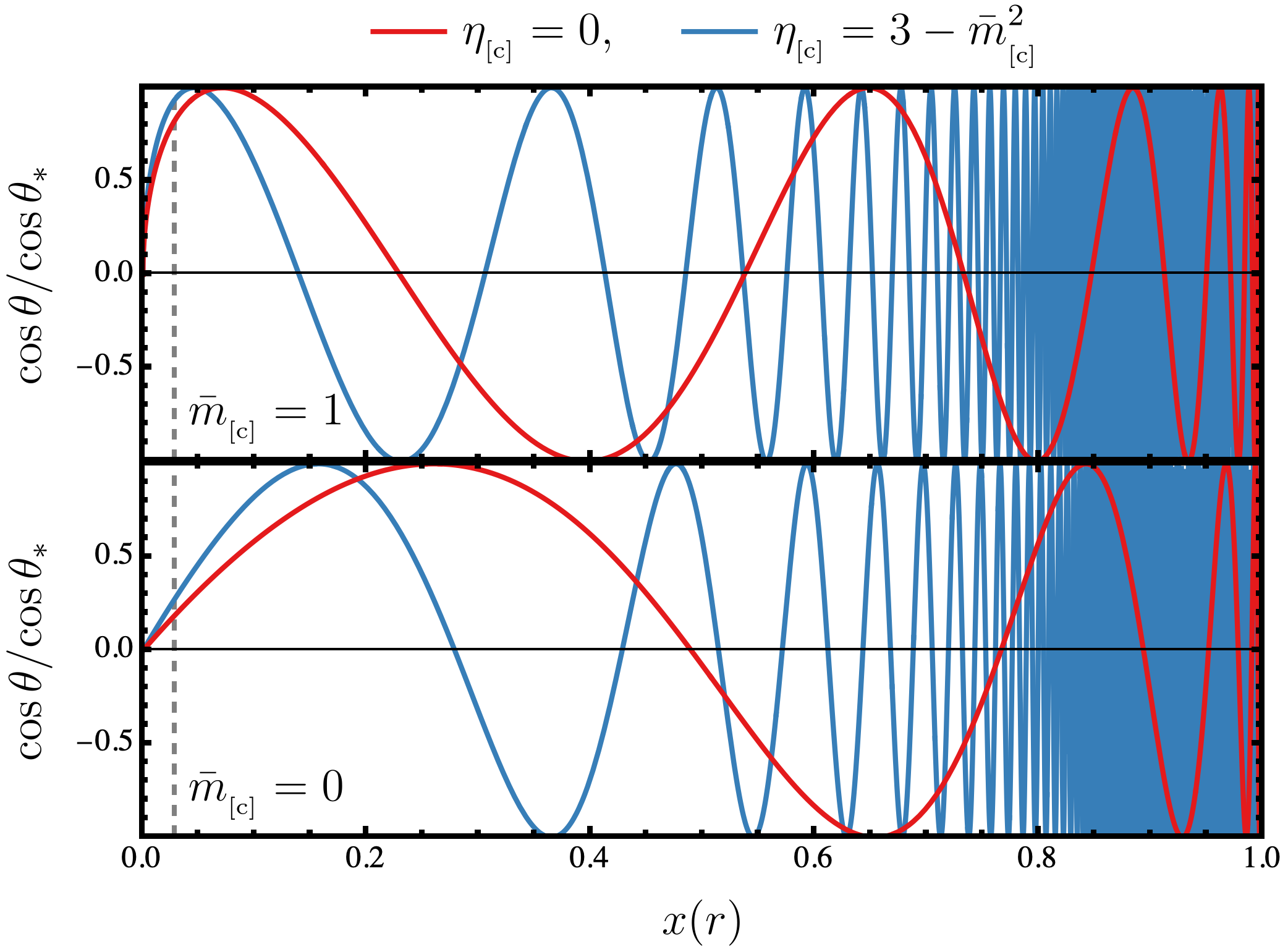}  \ & \ \includegraphics[width=.45\linewidth]{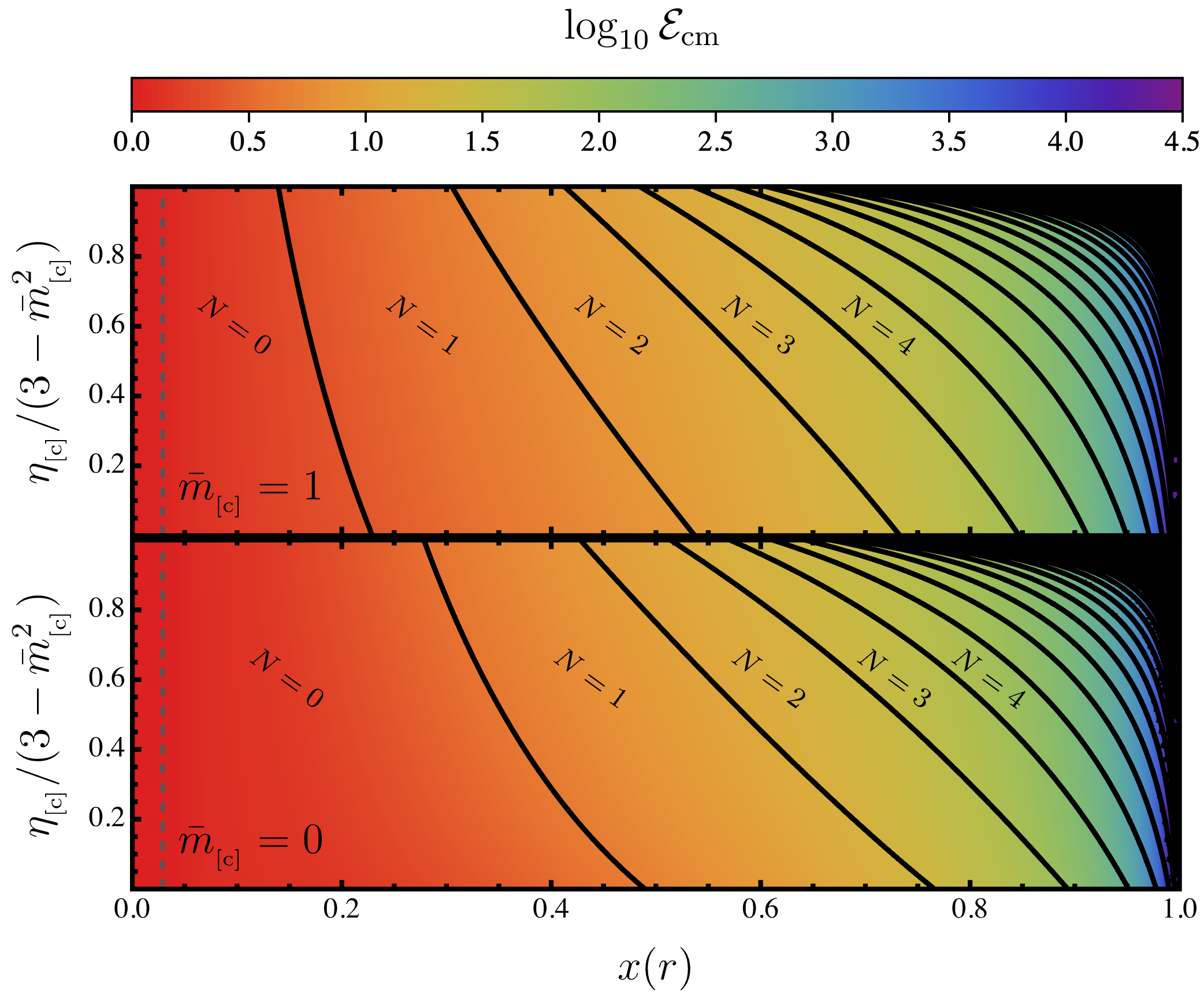}
    \end{tabular}}
    \caption{
    {\bf Left:} Libratory motion of critical particles originating at infinity on the equatorial plane and asymptotically spiraling onto spherical orbits at the horizon. 
    Trajectories of marginally bound (top row) and null (bottom row) particles with minimal (red) and maximal (blue) Carter constant are shown. The radius is shown using compactified variable \eqref{eq:RadComp} with the retrograde MBCO radius indicated by gray dashed lines.
    {\bf Right:} Enhancement factor for collisions between critical particles falling in from infinity and a retrograde disk particle, plotted as a function of Carter constant and collision radius. 
    The plot is partitioned by $N$, the minimal number of equatorial crossings of the critical particle before collision with the first five regions labeled. 
    }
    \label{fig:DiskCrossingS2}
\end{figure*}

In collision scenario 1, the MB/UB generic particle collides with a PD particle.
MB/UB generic particles reach the horizon in finite Mino time. 
Fig.~\ref{fig:GenericCross} shows the minimal number of equatorial crossings MB/UB particles undergo before reaching the horizon.
A sizable fraction of MB/UB generic particles can collide with the disk arbitrarily close to the horizon without any prior disk crossings, 
including particles with no equatorial crossings and particles with a single crossing occurring beyond the disk’s outer edge.
Thus, under collision scenario 1, the BH--PD is a viable astrophysical HEPC.

Collision scenarios 2 and 3 include a MB/UB critical particle. MB/UB critical particles take infinite Mino time to reach the horizon and undergo infinitely many equatorial crossings. 
In the left panel of Fig.~\ref{fig:DiskCrossingS2}, we plot the libratory behavior of critical particles initialized in the equatorial plane at infinity as a function of the radial compactification variable 
\begin{align}
    \label{eq:RadComp}
    x=1-\pa{1-\frac{M}{r}}^{1/4},
\end{align}
which maps infinity and the horizon to $x=0$ and $x=1$, respectively. In this mapping, the retrograde MSCO radius is at $x=1-(8/9)^{1/5}\approx0.029$. The radii $\hat r_{{\rm min,\max}}$ decrease monotonically with $\bar{m}_{\rm [c]}$ and $\eta_{\rm [c]}$, and are therefore minimized and maximized at $\cu{\bar{m}_{\rm [c]},\eta_{\rm [c]}}=\cu{0,0}$ and $\cu{1,1}$, respectively. Furthermore for finite $N$, $\hat r_{\rm min}-r_{\rm H}$ is always finite, so a collision including a critical particle that has undergone crossings $N$ cannot be brought arbitrarily close to the horizon. Thus, we examine the enhancement factor at fixed $N$ for collision scenarios 2 and 3~\footnote{In both scenarios, the enhancement factor is monotonic in all parameters.}.

In collision scenario 2, the MB/UB critical particle collides with an RD disk particle. 
The enhancement factor is minimized in the flat-space limit, where $\mathcal{E}_{\rm cm}\approx1$.
For fixed number of equatorial crossings of the critical particle $N$, the enhancement factor is minimized and maximized at $\cu{r,\bar{m}_{\rm [c]},\eta_{\rm [c]}}=\cu{\hat r_{\rm max},1,2}$ and $\cu{\hat r_{\rm min},0,0}$, respectively~\footnote{\label{ft:numerical}This statement is true within numerical error for $N
\in\cu{1,2,3,4}$ as an explicit form of $\hat r_{\rm min,max}$ is unknown.}.
When the critical particle has undergone $N$ crossings prior to collision, the enhancement factor is bounded by
\begin{align}
   N&=0:\quad {\color{white}0.0}0<\log_{10}\mathcal{E}_{\rm cm} \lesssim 0.65,\nonumber\\
   N&=1:\quad 0.27 \lesssim \log_{10}\mathcal{E}_{\rm cm} \lesssim 1.37,\nonumber\\
   N&=2:\quad 0.66 \lesssim \log_{10}\mathcal{E}_{\rm cm} \lesssim 2.06,\nonumber\\
   N&=3:\quad 0.89 \lesssim \log_{10}\mathcal{E}_{\rm cm} \lesssim 2.74,\nonumber\\
   N&=4:\quad 1.05 \lesssim \log_{10}\mathcal{E}_{\rm cm} \lesssim 3.42.\nonumber
\end{align}
Cross-sections of the enhancement factor are shown in the right panel of Fig.~\ref{fig:DiskCrossingS2}. 

\begin{figure}
    \centering
    \includegraphics[width=\linewidth]{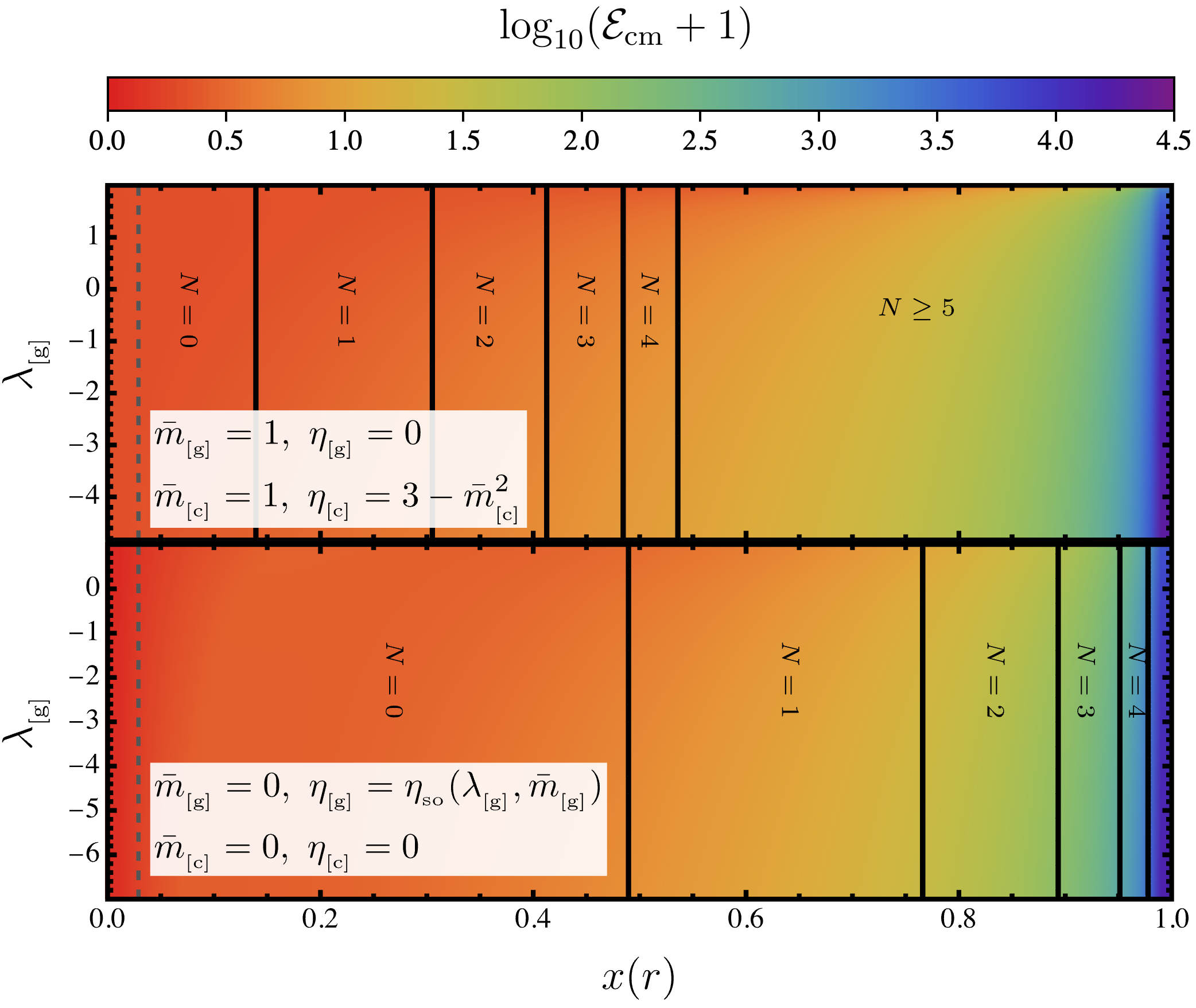}
    \caption{Enhancement factor for collisions between generic and critical particles falling in from infinity, delineated by $N$, the minimal number of equatorial crossings of the critical particle before collision. We explicitly label the first four such regions, as well as the region where $N\geq 5$.}
    \label{fig:DiskCrossingS3}
\end{figure}

In collision scenario 3, the MB/UB critical particle collides with a MB/UB generic particle. 
The enhancement factor is minimized in the flat-space limit when both particles are null, where $\mathcal{E}_{\rm cm}\approx0$.
For fixed number of equatorial crossings of the critical particle $N$ (relevant to scenario 3 when an accretion disk is present), the enhancement factor is minimized and maximized at $\cu{r,\bar{m}_{\rm [c]},\eta_{\rm [c]},\bar{m}_{\rm [g]},\eta_{\rm [g]},\lambda_{\rm [g]}}=\cu{\hat r_{\rm max},1,2,1,0,2}$ and $\cu{\hat r_{\rm min},0,0,0,0,-2\pa{1+\sqrt{2}}}$, respectively~${}^{\ref{ft:numerical}}$.
When the critical particle has undergone $N$ crossings prior to collision, the enhancement factor is bounded by
\begin{align}
   N&=0:\quad -\infty < \log_{10}\mathcal{E}_{\rm cm} \lesssim 0.73,\nonumber\\
   N&=1:\quad 0.22 \lesssim \log_{10}\mathcal{E}_{\rm cm} \lesssim 1.45,\nonumber\\
   N&=2:\quad 0.26 \lesssim \log_{10}\mathcal{E}_{\rm cm} \lesssim 2.13,\nonumber\\
   N&=3:\quad 0.28 \lesssim \log_{10}\mathcal{E}_{\rm cm} \lesssim 2.82,\nonumber\\
   N&=4:\quad 0.29 \lesssim \log_{10}\mathcal{E}_{\rm cm} \lesssim 3.50.\nonumber
\end{align}
Cross-sections of the enhancement factor are shown in Fig.~\ref{fig:DiskCrossingS3}.

For both collision scenarios 2 and 3, the CM energy enhancement at the first few equatorial crossings is modest. 
This suggests that a BH--RD is unlikely to serve as an astrophysical HEPC and that collision scenario 3 does not significantly contribute to the high-energy collisions near the BH--PD.

In the absence of a persistent accretion disk, collision scenario 3 can occur. 
Ideally, one would determine the minimal number of crossings for any pair of generic and critical particle trajectories that meet at the horizon.
A subset of such trajectories that intersect only at the horizon can be identified by selecting those that occupy disjoint polar-angle regions at all finite radii except the horizon.

\begin{figure}
    \centering
    \includegraphics[width=\linewidth]{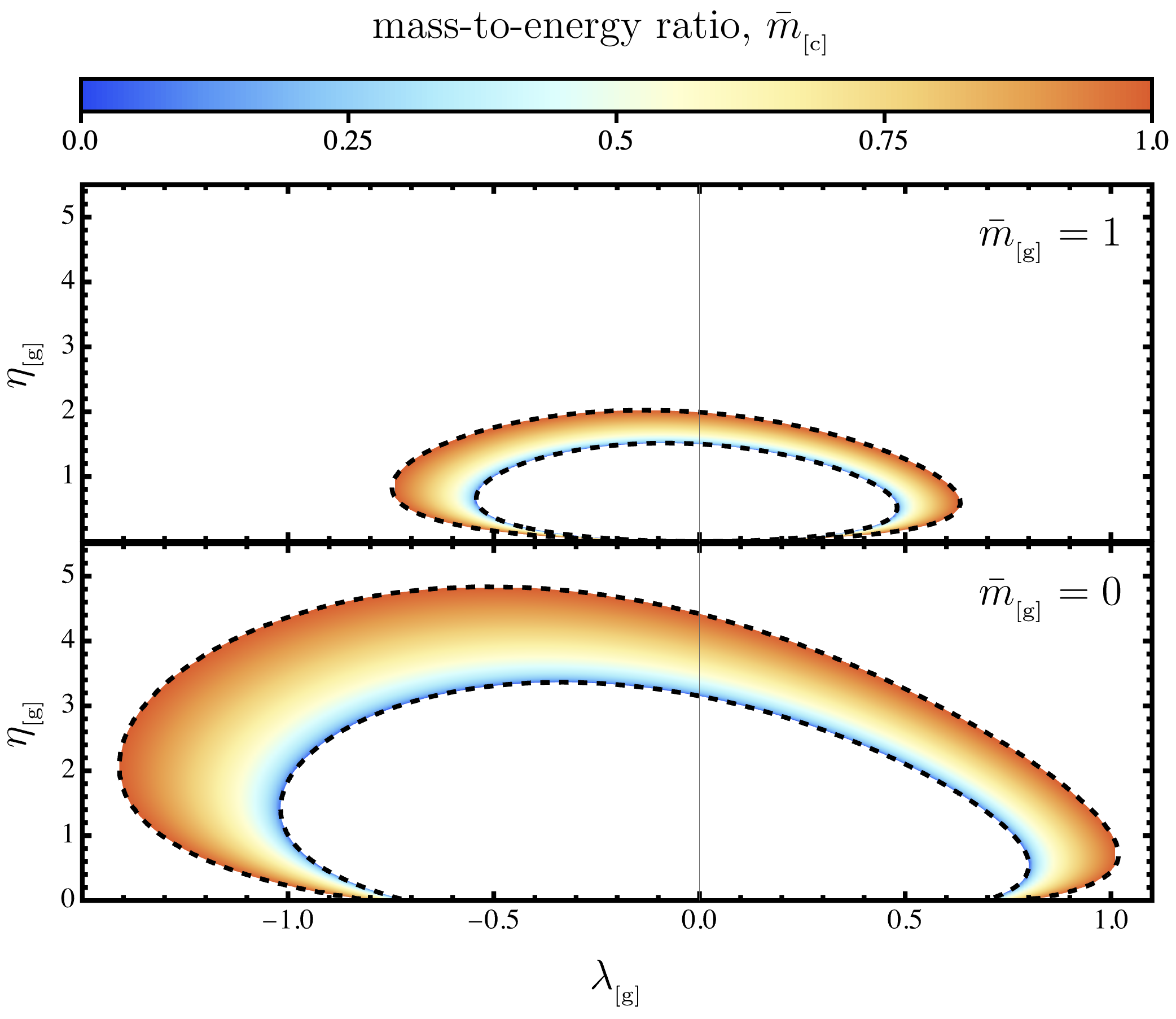}
    \caption{Boundary of phase-space region where a generic particle and a critical particle can occupy disjoint polar-angle regions at all finite radii, intersecting only at the horizon.
    The critical particle has maximal Carter constant $\eta_{\rm [c]}=3-\bar{m}_{\rm [c]}^2$ and the generic particle is marginally bound (top) or null (bottom).}
    \label{fig:ApexS3}
\end{figure}

Generic particles can librate over a larger region than the critical particles. Consequently, a subset of MB/UB generic particles can collide with a MB/UB critical particle at the horizon without entering the critical particle's libration region. 
That is, the generic and critical particle trajectories can be guaranteed to intersect only at the horizon (and possibly at infinity),
\begin{align}
\ab{\frac{\cos\theta_{{\rm [g]}}(r)}{\cos\theta_{*{\rm [c]}}}} \
    \begin{cases}
    \label{eq:ApexCollision}
        > 1, \quad & r\in(r_{\rm H},\infty)\\
        = 1, \quad & r=r_{\rm H}
    \end{cases},
\end{align}
where the subscripts indicate evaluation at the conserved quantities of the critical or generic particle.
In Fig.~\ref{fig:ApexS3}, we plot cross-sections of the boundary for the phase-space region where generic particles the condition~\eqref{eq:ApexCollision} can be satisfied~\footnote{Comparing phase-space cross-section of fixed $\cu{\bar{m}_{\rm [g]},\bar{m}{\rm [c]},\eta_{\rm [c]}}$ (i.e., $(\lambda_{\rm [g]},\eta_{\rm [g]})$-planes), the region where \eqref{eq:ApexCollision} can be satisfied grows as $\eta_{\rm [c]}\to0$ until it becomes the region where the particle can reach the horizon without crossing the equatorial place, $N_{\rm H}=0$. See Fig.~\ref{fig:GenericCross}.}. 
A sizable fraction of MB/UB generic-critical particle pairs can collide arbitrarily close to the horizon without their trajectories crossing prior.
Thus, under collision scenarios 3, the BH--ND is a viable astrophysical HEPC.

\emph{Discussion ---}
In this work, we have shown that scattering can moderate the CM energy of collisions between particles within an accretion flow and particles infalling from infinity near a BH, and have calculated the resulting CM energy bounds.
Furthermore, we disentangle special relativistic effects (the particle masses and energies) and the general relativistic phenomena underlying the BSW effect (the enhancement factor) in the CM energy.

When incorporating scattering, we find that the BH--RD is unlikely to constitute an astrophysical HEPC. 
In contrast, the BH--PD and the BH--ND remain viable candidates. 
The BH--PD is likely the more efficient of the two due to the persistence of the accretion disk.
We note that X-ray spectra observations support the existence of rapidly rotating BH--PD systems (e.g.,~\cite{Brenneman2013,Reynolds2021,Bambi2021,Draghis2023,Ricarte2025}); whereas, only a few observations are consistent with retrograde spins, which are far from extremal (e.g.,~\cite{Dauser2010,Steiner2012,Morningstar2014,Zdziarski2025}).
While this work provides an important step towards understanding the capacity of astrophysical BHs to act as HEPCs, 
there are many outstanding questions beyond the scope of this work. These include, but are not limited to, the following inquiries.

Firstly, ``What is the role of secondary collisions?''
An initial collision produces daughter particles that may subsequently undergo secondary collisions~\cite{Grib2010}, potentially increasing the high-energy collision rates and production efficiencies of BH systems. 
Near-horizon collisions involving a circular orbiter, which occur around the BH-PD, can readily produce critical daughter particles~\cite{Gates2023,Gates2025b}. 
For the BH--RD, it remains unclear whether a collision is likely to produce critical daughter particles which can reach closer to the BH horizon.

Secondly, ``What are the production rates and efficiencies of these collisions?''
Estimating the rate and efficiency of producing high-energy collisions might clarify how astrophysical UHPECs compare to terrestrial colliders. Calculating these quantities requires a prescription for the distribution of disk particles and infalling particles~\cite{Novikov1973,Potter2021}.

Lastly, ``Are these collisions observable?'' Combining collision rates with escape probabilities and expected energies of escaping particles can provide estimates of the number and energy fluxes of emission from high-energy collisions~\cite{Banados2011,Schnittman2014}. 
A formalism has been developed to examine the escape probability and expected energy of emission from a collision featuring a circular orbiter, which showed that such collisions produce reasonably observable emission~\cite{Gates2023}. It remains an open question whether the other collision scenarios outlined here also yield observable emission.

\emph{Acknowledgments ---}
DG thanks 
Alejandro C{\'a}rdenas-Avenda{\~n}o and Aaron Held for comments on this work. 
DG acknowledges financial support from the National Science Foundation (AST-2307887). 
This publication is funded in part by the Gordon and Betty Moore Foundation (Grant \#13526). 
It was also made possible through the support of a grant from the John Templeton Foundation (Grant \#63445). 
The opinions expressed in this publication are those of the author and do not necessarily reflect the views of these Foundations. 

\bibliography{BHColliderComparison.bib}

\end{document}